\documentclass[doublecol]{epl2}

\usepackage{amssymb}
\usepackage{amsmath}
\usepackage{epsf}
\usepackage{graphicx}
\usepackage{epsfig}

\title{Stability of a granular layer on an inclined ``fakir plane"}

\author{J. Benito\inst{1,2} \and Y. Bertho\inst{1} \and I. Ippolito\inst{3} \and P. Gondret\inst{1}}
\shortauthor{J. Benito, Y. Bertho, I. Ippolito and P. Gondret}

\institute{
  \inst{1} Univ Paris-Sud, Univ Paris 6, CNRS, Lab FAST, B\^at.~502, Campus Univ, F-91405 Orsay, France\\
  \inst{2} Instituto de F\'isica Aplicada, CONICET, Departamento de F\'isica, Universidad Nacional de San Luis, 5700 San Luis, Argentina\\
  \inst{3} Grupo de Medios Porosos, CONICET, Departamento de F\'isica, Facultad de Ingenier\'ia, Universidad de Buenos Aires, 1063 Buenos Aires, Argentina
}

\abstract{We present here experimental results on the effect of a forest of cylinder obstacles (nails) on the stability of a granular layer over a rough incline, in a so-called ``fakir plane" configuration. The nail forest is found to increase the stability of the layer, the more for the densest array, and such an effect is recovered by a simple model taking into account the additional friction force exerted by the pillar forest onto the granular layer.}

\begin{document}

\maketitle

\section{Introduction}

The flow and stability of granular layers on inclined planes have been widely studied in the last few years because it has a great interest from both practical and fundamental point of view \cite{GDRMidi04}. From a practical point of view, there is for instance an important need in risk management of rocks and snow avalanches for a better prediction of the stability and flow rules of granular media on rough inclines \cite{Mangeney03,Naaim03}. From a fundamental point of view, the transition from the ``solid" to the ``liquid" state of granular media is still an open question, and the inclined plane is a widely useful tool for investigating the unknown rheology of such complex matter \cite{GDRMidi04}. Numerous studies have been done concerning the stability of such a granular layer close to the flow transition \cite{Daerr99,Pouliquen99,Pouliquen02,Silbert01,Goujon03,Pouliquen04,Josserand06,Borzsonyi07,Borzsonyi08,Staron08}. A key result is that the stability of a layer of height $h$ on a rough incline of slope $\theta$ is characterised by two limiting curves in the ($h$, $\theta$) parameter space corresponding respectively to the spontaneous starting and stopping of the flow. Below the stopping curve, the granular layer is stable and no avalanche can develop. Above the starting curve, the layer is unstable and must flow. In between the two curves, the granular layer can be either in a ``liquid" (flowing) or a ``solid" (at rest) state, and avalanches can be triggered by small perturbations, that have to be larger when closer to the stopping curve \cite{Daerr99}. The key point is that $\tan\theta$, which corresponds to the ratio of the tangential to the normal forces, is related to the effective friction coefficient of the granular layer with the bottom plate. Thus exploring such stability and flow of a granular layer on an incline gives access to the effective granular friction forces \cite{Pouliquen02,Pouliquen08}. The two starting and stopping curves follow the same trend with a shift of only a few degrees in $\theta$. For large $h$ values, the two curves go asymptotically to the values corresponding to the two characteristic angles of stability of ``infinite" granular piles: the maximum angle of stability and the angle of repose. The typical shape of the two curves are qualitatively the same whatever the experimental configuration, such as the roughness of the plane (velvet clothe \cite{Daerr99}, sand paper \cite{Borzsonyi08}, glued beads \cite{Pouliquen99,Goujon03}) or the flowing granular material (spherical beads \cite{Daerr99,Pouliquen99,Goujon03,Borzsonyi08}, sand \cite{Pouliquen99,Borzsonyi08}, or even more anisotropic material such as copper \cite{Borzsonyi08}). These starting and stopping curves have been recovered by discrete numerical simulations with collisional and frictional forces \cite{Silbert01,Staron08} and different theoretical approaches based on physical arguments \cite{Josserand06,Wyart09}. The theoretical prediction \cite{Ertas02} for the link between the stopping height at a given angle $\theta$ and the correlation length between flowing grains have been put in light both experimentally \cite{Pouliquen04} and with discrete numerical simulations \cite{Staron08}.
In this Letter, we investigate the influence of a forest of cylindrical obstacles, experimentally a forest of nails, on the stability of a granular layer onto a so-called ``fakir plane". We present first experimental results and then a model that allows one to recover the essential of the experimental observations. The findings may be of importance in risk management for the possible stabilisation of snow avalanches or rocks for example by a forest of trees.

\section{Experimental setup}

\begin{figure}[t!]
\includegraphics[width=\linewidth]{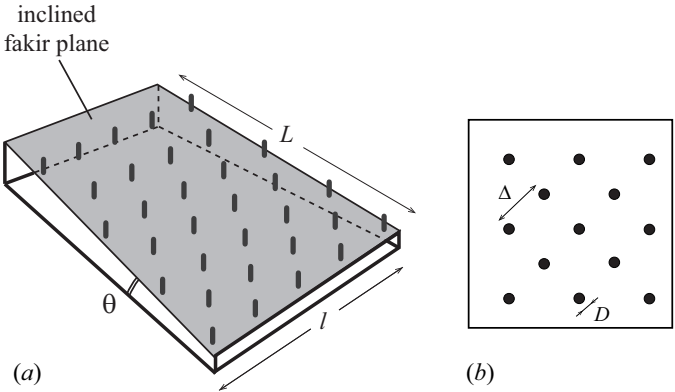}
\caption{($a$)~Sketch of the experimental setup and notations used. ($b$)~Sketch of the fakir plane (top view).}
\label{Fig01}
\end{figure}

The experimental setup consists of a plane that can be inclined by an angle $\theta$ from the horizontal up to about $45$ degrees [fig.~\ref{Fig01}($a$)]. The plane is covered by a velvet clothe which is chosen so that the grains have a larger friction with it than between themselves. Nails of diameter $D=2$\,mm and length $H=35$\,mm are hammered in the plane (perpendicular and tips up) along a square lattice with a distance $\Delta$ between the nearest neighbours as shown in fig.~\ref{Fig01}($b$), leading to a so-called ``fakir plane". Two different fakir planes have been built with $\Delta=10\sqrt{2}$\,mm $\simeq 14$\,mm and $\Delta=5\sqrt{2}$\,mm $\simeq 7$\,mm ($\Delta/D = 7$ and 3.5). The particles used in the experiments are sieved glass beads (density $\rho\simeq 2500$\,kg m$^{-3}$) of diameter $d_g$ ranging from 0.13\,mm to 1.2\,mm with a relative dispersion in diameter of order 0.1. The corresponding diameter values have been measured with a particle analyser (Morphologi G3, Malvern). The dimensions of the fakir planes used in the present study, with a length $L=300$\,mm and width $l=180$\,mm, are significantly smaller than the ones used in previous studies \cite{Daerr99,Pouliquen99,Goujon03,Borzsonyi07,Borzsonyi08}. Nevertheless, we have checked this does not change significantly the results in the ``non-fakir" case ($\Delta=\infty$) and believe that the results would not be affected significantly with larger dimensions in the ``fakir" cases. It is worth noting that the granular layer deposit on the plane is never higher than the nails so that the height $H$ of the nails can be considered as infinite and does not play any role here. Note also that the ``fakir plane" just refers to the used plane hammered with nails but does not refer to any possible ``fakir" state of materials laying at the top of nails rather than on the bottom of the plate as exhibited by liquids drops in an induced ``hydrophobic" state by a forest of microscopic pillars \cite{Quere02}. The different experiments presented here correspond to the following ranges for the cylinder/grain size ratio $D/d_g$ and the number of grains between two nails $(\Delta - D)/d_g$: $1.5 \lesssim D/d_g \lesssim 15$ and $4 \lesssim (\Delta - D)/d_g \lesssim 100$.

The experimental procedure consists in pouring a known amount of grains onto the horizontal plane of surface $S = L \times l$, and then slowly inclining it to trig successive avalanches that affect the entire surface, following the procedure described in \cite{Daerr99}. The weighting of the grains that have flowed out of the plane after each avalanche at an angle $\theta$ allows one to infer the thickness $h$ of the remaining layer assuming a constant solid fraction $\phi$ of the grain layer ($\phi \simeq 0.6$). This procedure allows to reconstruct the two curves ($h$, $\theta$) that characterise the stability of the granular layer on the inclined plane: the starting curve at which the layer starts flowing and the stopping curve at which the layer stops flowing.

\section{Experimental Results}

\subsection{Inclined plane without pillars}

\begin{figure}[t]
\includegraphics[width=\linewidth]{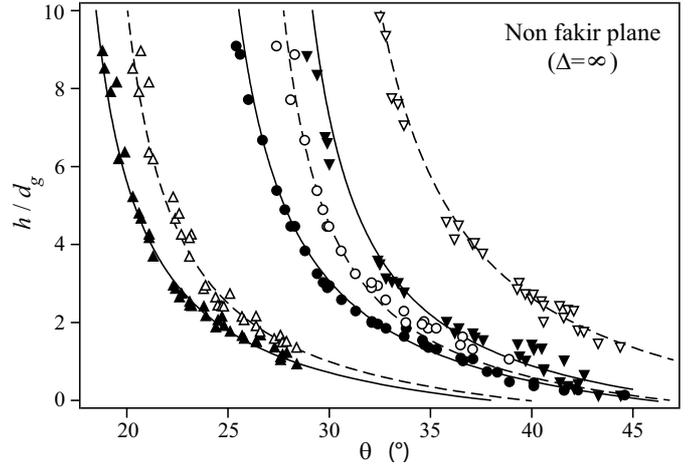}
\caption{Dimensionless layer thickness $h/d_g$ as a function of the angle $\theta$ for the non-fakir plane and different grain sizes $d_g$ with open (solid) symbols for the starting (stopping) curves: ($\triangledown, \blacktriangledown$)~$d_g=0.33$\,mm, ($\circ, \bullet$)~$d_g=0.75$\,mm and ($\vartriangle, \blacktriangle$) $d_g=1.2$\,mm. (--~--)~Best fits of starting data by eq.~(\ref{Eq:hyp}) with fitting parameter values ($\triangledown$)~$\theta_{1m} \simeq 27.6^{\circ}$, $\theta_{2m}\simeq 59.1^{\circ}$ and $\lambda_m \simeq 0.31$\,mm, ($\circ$)~$\theta_{1m} \simeq 25.4^{\circ}$, $\theta_{2m} \simeq 46.9^{\circ}$ and $\lambda_m \simeq 0.71$\,mm, ($\vartriangle$)~$\theta_{1m} \simeq 19.1^{\circ}$, $\theta_{2m} \simeq 36.4^{\circ}$ and $\lambda_m \simeq 1.1$\,mm. (---)~Best fits of stopping data by eq.~(\ref{Eq:hyp}) with the fitting parameters: ($\blacktriangledown$)~$\theta_{1r} \simeq 23.1^{\circ}$, $\theta_{2r} \simeq 58.8^{\circ}$ and $\lambda_r \simeq 0.31$\,mm, ($\bullet$)~$\theta_{1r} \simeq 22.7^{\circ}$, $\theta_{2r} \simeq 46.3^{\circ}$ and $\lambda_r \simeq 0.79$\,mm, ($\blacktriangle$)~$\theta_{1r} \simeq 17.3^{\circ}$, $\theta_{2r} \simeq 35.5^{\circ}$ and $\lambda_r \simeq 1.1$\,mm.}
\label{Fig02}
\end{figure}

\begin{figure}[t]
\includegraphics[width=\linewidth]{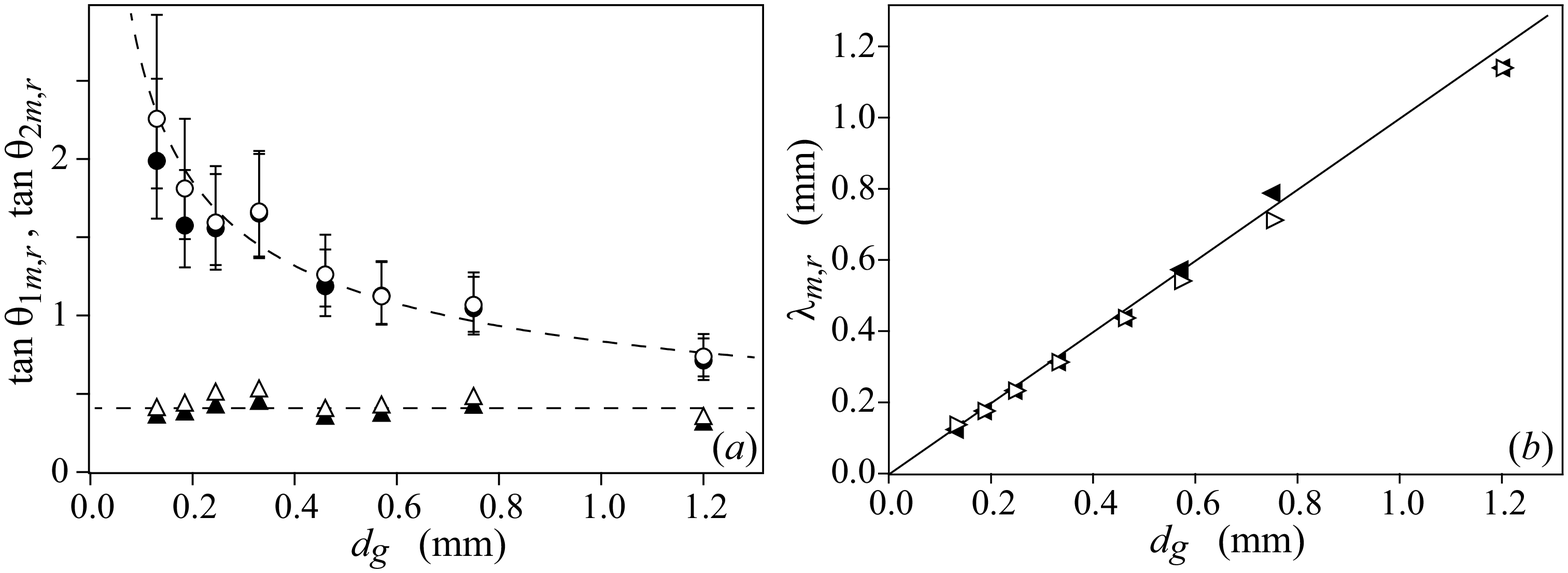}
\caption{Fitting parameters of eq.~(\ref{Eq:hyp}): ($a$) $\tan \theta_{1m,r}$ ($\vartriangle, \blacktriangle$) and $\tan \theta_{2m,r}$ ($\circ, \bullet$), and ($b$) $\lambda_{m,r}$ as a function of the grain diameter $d_g$ for starting (open symbols) and stopping (solid symbols) cases. (-~-~-)~Guidelines for the eyes. (---)~Linear fit $\lambda_{m,r}=d_g$.}
\label{Fig03}
\end{figure}

Let us first present our measurements for the ``non-fakir" case ($\Delta=\infty$). Figure~\ref{Fig02} shows the starting and stopping curves in the ($h$, $\theta$) parameter space obtained for different grain diameters. These curves can be fitted by the following equation where the indexes $m$ (for ``motion") and $r$ (for ``repose") refer to the starting and stopping cases respectively:
\begin{equation}
\tan\theta_{m,r}(h)= \tan\theta_{1m,r} + \frac{\tan\theta_{2m,r} - \tan\theta_{1m,r}}{1+h/\lambda_{m,r}}.
\label{Eq:hyp}
\end{equation}
In this equation $\theta_{1m,r}$ corresponds to the characteristic starting or stopping angle of the granular pile (layer of infinite height $h\rightarrow \infty$), $\theta_{2m,r}$ corresponds to the characteristic starting or stopping angle for vanishing layer thickness $h$ ($h\rightarrow 0$), and $\lambda_{m,r}$ is the characteristic height of influence of the bottom rough condition for the starting or stopping case respectively. The indexes $m,r$ will be dropped in the following to avoid too heavy notation but one have to keep in mind that $\theta_1$, $\theta_2$ and $\lambda$ are not the same for the starting and stopping cases. Such an equation was already used by \cite{Pouliquen02,Borzsonyi07} for example, but an exponential dependence $\exp(-h/\lambda$) instead of the $(1+ h/\lambda)^{-1}$ dependence has also been used by \cite{Daerr99,Pouliquen99,Goujon03,Staron08}. Both functional dependences coincide at low $h/\lambda$ values, giving a linear decrease of $\tan\theta$ as a function of $h$, but differ in the asymptotic relaxation of $\theta$ towards $\theta_1$ for large $h$ values. The fits of our data by eq.~(\ref{Eq:hyp}) appearing in solid and dashed lines in fig.~\ref{Fig02} for the stopping and starting curves respectively are quite close to the experimental data. The fitting values of the three parameters $\tan\theta_1$, $\tan\theta_2$ and $\lambda$ are shown in fig.~\ref{Fig03} for all the used grain sizes $d_g$ from 0.13 to 1.2 mm. The $\tan \theta_1$ value lies in the range $0.3 \lesssim \tan \theta_1 \lesssim 0.5$ ($17^{\circ} \lesssim \theta_1 \lesssim 28^{\circ}$) with the starting value $\theta_{1m}$ above the stopping one $\theta_{1r}$ by about 2 degrees. No systematic variation is observed for $\tan\theta_1$ with $d_g$. By contrast, the $\tan \theta_2$ value decreases significantly with increasing $d_g$ from about 2 down to 0.7 ($35^{\circ} \lesssim \theta_2 \lesssim 65^{\circ}$) with the starting value $\theta_{2m}$ above the stopping one $\theta_{2r}$ by also about 2 degrees. This monotonic dependence of $\theta_2$ found here for glass beads on velvet clothe is different from the non-monotonic dependence observed by \cite{Goujon03} for glass beads on a plane with glued beads of a given size. The difference between $\tan \theta_2$ and $\tan \theta_1$ here decreases which means that the effective friction coefficient between the grains and the rough plane is less different than the effective friction coefficient between the grains when the grain size is larger. The parameter $\lambda_{m,r}$ is found to increase linearly with the grain diameter $d_g$ with no significant difference between $\lambda_m$ (starting case) and $\lambda_r$ (stopping case) values. Note that the fitting value $\lambda_{m,r}/d_g \simeq 1$ we find here for our case of glass beads can be significantly different for non-spherical particles \cite{Borzsonyi08}.

\subsection{Inclined plane with pillars}

\begin{figure}[t]
\includegraphics[width=\linewidth]{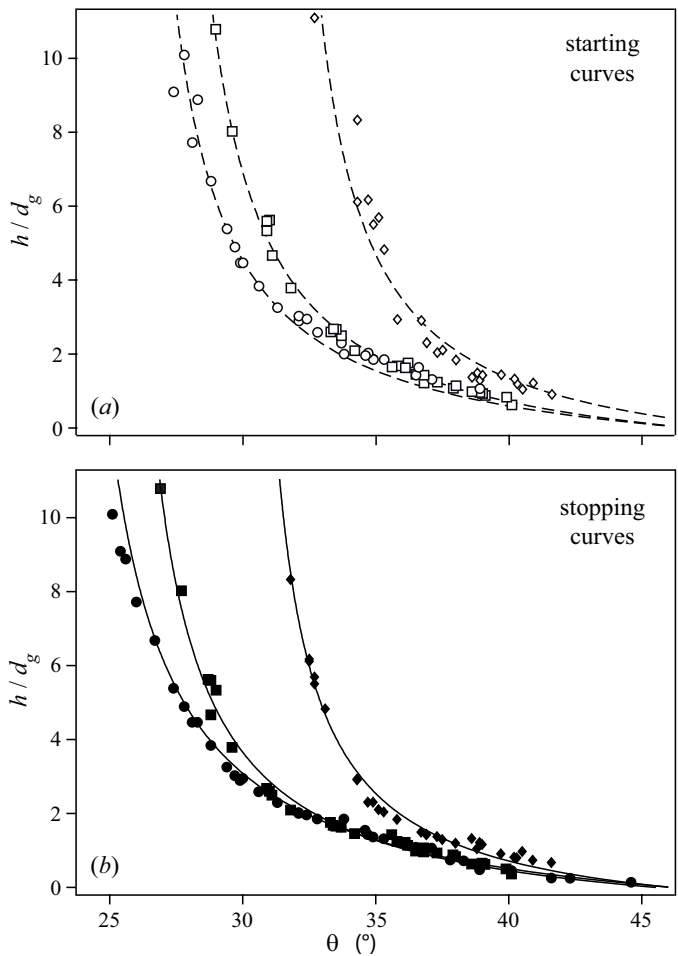}
\caption{Dimensionless layer thickness $h/d_g$ for ($a$) starting and ($b$) stopping cases as a function of the angle $\theta$ for grains of diameter $d_g=0.75$\,mm and for ($\circ, \bullet$) $\Delta=\infty$, ($\square, \blacksquare$) $\Delta=14$\,mm and ($\lozenge, \blacklozenge$) $\Delta=7$\,mm. (-- --)~Best fits of starting data by eq.~(\ref{Eq:hyp}) with the fitting parameters: $\lambda_m \simeq d_g$, $\theta_{2m} \simeq 46.9^\circ$ and ($\circ$)~$\theta_{1m}^\infty \simeq 25.4^\circ$, ($\square$)~$\theta_{1m} \simeq 26.7^{\circ}$, ($\lozenge$)~$\theta_{1m} \simeq 31.1^\circ$. (---)~Best fits of stopping data by eq.~(\ref{Eq:hyp}) with fitting parameters values $\lambda_r \simeq d_g$, $\theta_{2r} \simeq 46.3^{\circ}$ and ($\bullet$)~$\theta_{1r}^\infty \simeq 22.7^\circ$, ($\blacksquare$)~$\theta_{1r} \simeq 24.8^\circ$, ($\blacklozenge$)~$\theta_{1r} \simeq 29.8^\circ$.}
\label{Fig04}
\end{figure}

\begin{figure}[t]
\includegraphics[width=\linewidth]{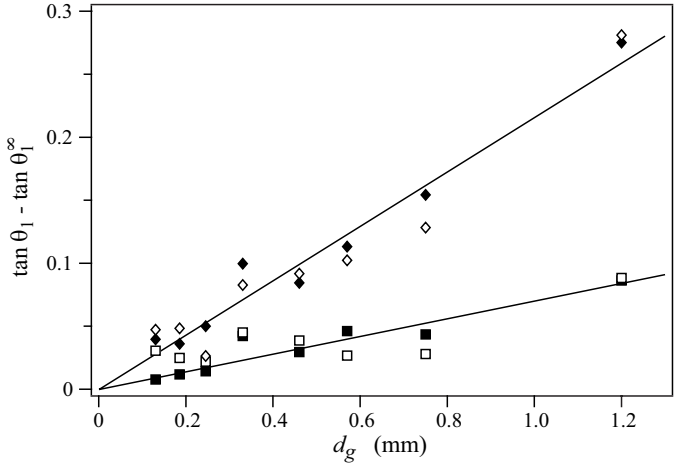}
\caption{Evolution of $\tan\theta_1$ relative to its value without pillars $\tan\theta_1^\infty$ ($\Delta=\infty$) as a function of the grain diameter $d_g$ for the two fakir planes with nail spacing $\Delta\simeq 14$\,mm ($\square, \blacksquare$) and 7\,mm ($\lozenge, \blacklozenge$). Open and solid symbols are related respectively to the starting and stopping cases. (---)~Linear fits with slopes 0.07\,mm$^{-1}$ for $\Delta\simeq 14$\,mm and 0.22\,mm$^{-1}$ for $\Delta\simeq 7$\,mm.}
\label{Fig05}
\end{figure}

The influence of a forest of cylindrical obstacles (pillars) on the starting and stopping curves has been studied by performing similar experiments on the fakir planes with two nail spacings $\Delta$ and different grain sizes $d_g$. Figure~\ref{Fig04} shows the typical effect of the nail forest on the ({\it a}) starting and ({\it b}) stopping curves for grains of diameter $d_g=0.75$\,mm.
The ($h$, $\theta$) curves obtained for the fakir planes present the same trend evolution as the non-fakir plane. Nevertheless, for thick enough layers at small enough angles, a significant shift of the curves is observed: for a given layer thickness $h$, a higher $\theta$ value is measured for the fakir case with a greater shift for the denser network (smaller spacing $\Delta$); and for a given angle $\theta$, a thicker layer $h$ is measured for a smaller $\Delta$. This clearly shows the stabilising effect of the nail forest. This stabilising effect vanishes for vanishing layer thickness at large angles where the different curves collapse.
In order to quantify the influence of the nail forest on the layer stability, we fit the different curves for the different planes with the same eq.~(\ref{Eq:hyp}) by keeping about constant the two parameters $\theta_2$ and $\lambda$ that are related to the influence of the bottom wall and already extracted in the case without obstacles. The obtained fitting curves, shown in fig.~\ref{Fig04} by dashed ({\it a}) or solid ({\it b}) lines for the starting ({\it a}) or stopping ({\it b}) data, with the only one remaining free fitting parameter $\theta_1$ are quite good. The effect of the pillar forest on the layer stability thus reduces to the modification of the $\theta_1$ value. The dependence of $\tan \theta_1$ for the starting and stopping cases relative to its value $\tan\theta_1^\infty$ without any pillars (at $\Delta=\infty$) is plotted in fig.~\ref{Fig05} for the different grain diameters $d_g$ and the two pillar networks $\Delta$. One can see that the difference $\tan\theta_1(\Delta)-\tan\theta_1^\infty$ is proportional to $d_g$ with a slope that increases for decreasing $\Delta$, with no significant differences between the starting (open symbols) and stopping (solid symbols) cases.

\section{Modelling}

\begin{figure}[t]
\includegraphics[width=\linewidth]{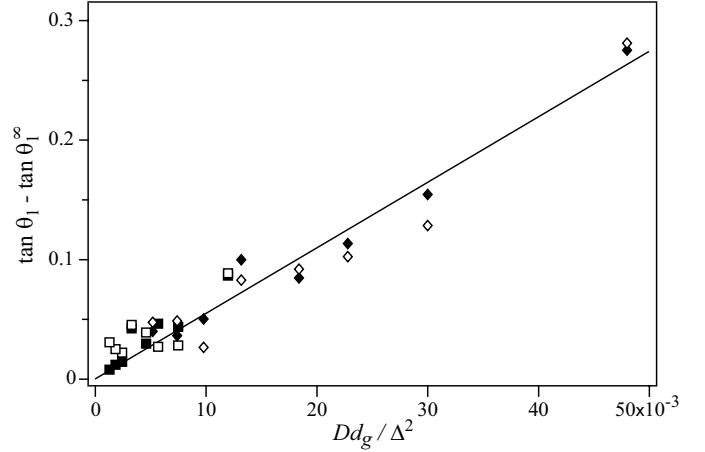}
\caption{Same data as in fig.~\ref{Fig05} but as a function of the ratio $Dd_g/\Delta^2$. (---)~Linear fit with slope 5.5.}
\label{Fig06}
\end{figure}

Let us now model the stabilisation played by the forest of pillars. It is known that a granular flow around one cylinder of diameter $D$ induces a drag force $F_c$ on this cylinder \cite{Albert99,Seguin11}. For a quasi-stationary regime, the force exerted by the cylinder forest on the granular flow should be opposite to the force exerted by the granular flow on the cylinder forest, and we shall consider it to be equal to the one-cylinder force times the number of cylinders by neglecting any strong flow interaction between the cylinders. The stress field of the grain contact network around a cylinder in relative motion has not been yet investigated, but the granular flow perturbation has been shown to be localised in a radial extension zone of about one cylinder diameter whatever the grain size \cite{Seguin11}.  The one-cylinder force $F_c$ is known to be $F_c=\alpha\rho\phi g D h_s^2$ independent of velocity, where $\rho\phi$ is the effective density of the granular medium ($\phi$ refers to the grain packing fraction), $h_s$ is the vertical penetration depth of the cylinder from the granular free surface, and $\alpha$ is a dimensionless coefficient depending on geometrical shape, and on friction and restitution coefficients of the grain/cylinder couple \cite{Albert99}. Note that some weak dependence of $F_c$ with the grain diameter $d_g$ (or the size ratio $d_g/D$) has been observed by some authors \cite{Albert99,Seguin11}, which will here be neglected in a first approximation. As for a forest of cylinders in a square array of spacing $\Delta$ the number of cylinders per unit area is $1/\Delta^2$, the force per unit surface exerted by the cylinder forest onto the granular layer is $F_{cf}=F_c/\Delta^2=\alpha \rho\phi g D h_s^2/\Delta^2$.
If one now considers the stability of a layer of thickness $h_s$ close to the surface of a granular pile and far away from any bottom, the stability of such a layer of length $L$ and width $l$ when it just starts or stops at the characteristic starting or stopping angle $\theta_1$ can thus be written as
\begin{equation}
\begin{split}
0=\rho\phi gh_s lL\sin\theta_1 - \mu_g \rho\phi g h_s lL \cos\theta_1\\ - \frac{lL}{\Delta^2}\alpha\rho\phi gD h_s^2 \cos\theta_1,
\end{split}
\label{eq2}
\end{equation}
where $\mu_g = \tan\theta_1^\infty$ corresponds to the internal friction of the granular packing and thus to the characteristic angle of the pile for the starting or stopping avalanche flow referred usually as the ``maximum angle of stability" ($\theta_{1m}^\infty$) and ``angle of repose" ($\theta_{1r}^\infty$) of the pile.
Equation~(\ref{eq2}) leads thus simply to
\begin{equation}
\tan\theta_1(\Delta)=\tan\theta_1^\infty +\alpha \frac{Dh_s}{\Delta^2}.
\label{eq3}
\end{equation}
The additional friction term [second term in the right hand side of eq.~(\ref{eq3})] coming from the cylinder forest increases thus the characteristic starting and stopping pile angles $\theta_{1m,r}$.

If one now considers the configuration of a granular layer of thickness $h$ on a fakir plane, its stability can thus be written as
\begin{equation}
\tan\theta(h,\Delta)= \tan\theta_1(\Delta) + \frac{\tan\theta_2 - \tan\theta_1(\Delta)}{1+h/\lambda}.
\label{eq4}
\end{equation}
This equation is the same as eq.~(\ref{Eq:hyp}) but with the effect of cylinder forest encoded here in the $\tan\theta_1$ term which is a function of the cylinder spacing $\Delta$ (and diameter $D$) given by eq.~(\ref{eq3}). The experimental results plotted in fig.~\ref{Fig05} shows that $\tan\theta_1-\tan\theta_{1}^{\infty}$ scales with $d_g$ and thus the additional friction term in eq.~(\ref{eq3}) scales with $d_g$. This means that the characteristic flowing thickness $h_s$ considered in model equation (3) scales with $d_g$ and does not correspond to the entire layer height $h$. By considering thus that $h_s\propto d_g$, model equation~(\ref{eq3}) predicts that the difference $\tan\theta_1-\tan\theta_{1}^{\infty}$ is proportional to the size ratio $Dd_g/\Delta^2$. The plot of our experimental data corresponding to $\tan\theta_1-\tan\theta_{1}^{\infty}$ as a function of $Dd_g/\Delta^2$ in fig.~\ref{Fig06} appears about linear with the slope 5.5. This slope value gives thus $\alpha h_s \simeq 5.5d_g$ which gives in turn $h_s \simeq d_g$ when considering the value $\alpha \simeq 5 \pm 1$ measured in \cite{Albert99}. The fact that the found characteristic flowing thickness $h_s$ scales with the grain diameter $d_g$ is not very surprising as granular matter is known to exhibit complex fluid behaviour with flow localisation \cite{GDRMidi04}. In particular flows are very localised in a thin surface layer as already visualised by \cite{Courrech05} for unsteady ``spontaneous" avalanches or by \cite{Crassous08} and references herein for forced flows: in the first case, the velocity profile exhibits an exponential decrease in the depth with a characteristic flowing layer that corresponds to about $(2\pm1)d_g$ \cite{Courrech05} ; in the second case, the linear part of the velocity profile related to the high granular flow is followed by an exponential decreasing part with a characteristic thickness $(1.1\pm0.1)d_g$ \cite{Crassous08}. Besides, the typical flowing thickness deduced from the modelling of the pile stabilisation by close lateral walls have also been shown to be of about $(2\pm1)d_g$ for the range of grain diameter used in the present study where the grain size is not too small to be hardly influenced by van der Waals interaction forces \cite{Courrech03,CourrechPhD03}. It is worth noting that $h_s$ is found here similar to $\lambda$, as $\lambda \simeq d_g$ and $h_s \simeq d_g$. This means that the effective friction and thus the stability of a granular layer on a rough plane may be related to the avalanche flow localisation. The granular flow may reach the bottom plane for low enough $h$ and not for high enough $h$. This may explain the shape of the $h(\theta$) curves of figs. 2 and 4 where $\tan\theta$ varies strongly at low $h/d_g$ and weakly at large $h/d_g$. Note that this would imply that the angle of the surface granular layer $\theta_s$ may be different from the plane angle for large enough granular layer $h/d_g$ as the granular flow does not ``feel" the bottom anymore: the granular layer would be no more homogeneous along the plane but would exhibit a streamwise negative gradient $\partial h/\partial x=-\tan(\theta_s-\theta)$. In the present study, we restrict to $h/d_g \lesssim 20$, where no significant thickness gradient is observed along the plane. For very large $h/d_g$ values ($h/d_g \gtrsim 20$), we indeed observe such a thickness gradient. In such a case, we believe that eq. (4) still holds but with the angle of the surface layer $\theta_s$ instead of the plane angle $\theta$. Such a regime have already be seen between two close lateral walls for which the pile angle is stabilised by the additional friction forces at the walls \cite{Courrech03} and may lead to the so-called ``super stable heap" for a large enough forced flow \cite{Taberlet03}. The effect of the pillar forest on the pile stabilisation appears to be very similar to the effect of close lateral walls, both cases leading to an effective additional frictional term.
Note that in the explored range of parameters, we always observed an angle $\theta_1$ smaller than $\theta_2$ but the contrary may be imagined for a high enough value of the additional friction term due to the pillar forest in eq.~(3).
Note also that when $h$ is smaller than $h_s$, the additional stabilising term of the cylinder forest in eq.~(3) is thus overestimated, but this term is negligible as the stabilising effect of the bottom is much larger and governs the layer stability in such a case. Moreover, since the measured drag force on a cylinder has been reported by \cite{Albert99,Seguin11} to increase weakly with decreasing grain size $d_g$ (or with the size ratio $d_g/D$), the variation of $\tan\theta_1(\Delta)-\tan\theta_{1}$ with $d_g$ and also $Dd_g/\Delta^2$ would be perhaps weakly sublinear.

The critical granular thickness necessary to observe a significant increase of its stability by a given pillar network can be estimated by the cross-over between the regime dominated by the bottom friction given by the second term of the right hand side of eq.~(\ref{eq4}) and the regime dominated by pillar forest friction given by the second term of the right hand side of eq.~(\ref{eq3}). As the variations of $\tan\theta_1$ are small compared to the difference $\tan\theta_2 - \tan\theta_1$, this critical height value $h_c$ is given by
\begin{equation}
\frac{\tan\theta_2 - \tan\theta_1}{1+h_c/\lambda} \sim \alpha \frac{Dh_s}{\Delta^2}.
\end{equation}
As this happens typically when $h/\lambda \gtrsim 1$, and as $\lambda \sim d_g$ and $h_s \sim d_g$, the previous equation leads approximately to
\begin{equation}
\frac{h_c}{d_g} \sim \frac{\tan\theta_2 - \tan\theta_1}{\alpha}\frac{\Delta^2}{Dd_g}.
\end{equation}
This predicts that the critical dimensionless layer thickness $h_c/d_g$ scales linearly with the dimensionless size ratio $\Delta^2/(Dd_g)$ for a constant ratio $(\tan\theta_2 - \tan\theta_1)/\alpha$, and thus that the critical layer thickness $h_c$ scales linearly with $\Delta^2/D$. Note that $(\tan\theta_2 - \tan\theta_1)/\alpha$ should not be constant but vary only weakly with $d_g$ as both $\tan\theta_2 - \tan\theta_1$ and $\alpha$ decrease with increasing $d_g$.

In this paper, we show a clear stabilising effect of an array of pillars for a granular layer on a rough incline. The experimental results are well fitted by a model equation taking into account the additional friction arising from the pillar forest when compared to the usual bottom friction on the rough incline. The key parameter appears to be the dimensionless size ratio $\Delta^2/(Dd_g)$. Additional experiments would be necessary to verify the dependence on the pillar diameter $D$ and the model prediction that the real flowing thickness $h_s$ should be smaller than the layer thickness $h$ for high enough $h$ and should scale with the grain diameter $d_g$. In forthcoming experiments, we intend to look at continuous feeding flows where stationary regimes may be reached. Finally, the present study may be useful for risk management as it predicts that a forest of trees characterised by a typical spacing $\Delta$ and diameter $D$ may stabilise some layer of discrete materials such as rocks or even snow on a slope. It is also related to the possible stabilisation of soil slopes by vegetation and its root system.

\acknowledgments
This work is supported by the French ANR project STABINGRAM No. 2010-BLAN-0927-01 and the CNRS-Conicet International Associated Laboratory LIA-PMF (Physics and Mechanics of Fluids). We are grateful to Bernard Perrin for the modelling.



\begin{thebibliography}{5}

\bibitem{GDRMidi04}
    \Name{G.D.R Midi}
    \REVIEW{Eur. Phys. J. E} {14}{2004}{341}.

\bibitem{Mangeney03}
    \Name{Mangeney-Castelnau, A. and Vilotte, J. -P. and Bristeau, M. O. and Perthame, B. and Bouchut, F. and Simeoni, C. \and Yerneni, S.}
    \REVIEW{J. Geophys. Res.} {108}{2003}{2527}.

 \bibitem{Naaim03}
    \Name{Naaim M., Faug T. \and Naaim-Bouvet F.}
    \REVIEW{Surv. Geophys.} {24}{2003}{569}.

\bibitem{Pouliquen96}
    \Name{Pouliquen O. \and Renaut N.}
    \REVIEW{J. Phys. II France}{6}{1996}{923}.

\bibitem{Daerr99}
    \Name{Daerr A. \and Douady S.}
    \REVIEW{Nature}{399}{1999}{241}.

\bibitem{Pouliquen99}
    \Name{Pouliquen O.}
    \REVIEW{Phys. Fluids}{11}{1999}{542}.

\bibitem{Pouliquen02}
    \Name{Pouliquen O. \and Forterre Y.}
    \REVIEW{J. Fluid Mech.}{453}{2002}{133}.

\bibitem{Silbert01}
    \Name{Silbert L. E., Ertas D., Grest G. S., Halsey T. C., Levine D. \and Plimpton S. J.}
    \REVIEW{Phys. Rev. E}{64}{2001}{051302}.

\bibitem{Goujon03}
    \Name{Goujon C., Thomas N., \and Dalloz-Dubrujeaud B.}
    \REVIEW{Eur. Phys. J. E} {11}{2003}{147}.

\bibitem{Pouliquen04}
    \Name{Pouliquen O.}
    \REVIEW{Phys. Rev. Lett.}{93}{2004}{248001}.

\bibitem{Josserand06}
    \Name{Josserand C., Lagr\'ee P.-Y. \and Lhuillier D.}
    \REVIEW{Europhys. Lett.}{73}{2006}{363}.

\bibitem{Borzsonyi07}
    \Name{B\"orzs\"onyi T. \and Ecke R.E}
    \REVIEW{Phys. Rev. E}{76}{2007}{031301}.

\bibitem{Borzsonyi08}
    \Name{B\"orzs\"onyi T., Halsey T.C. \and Ecke R.E}
    \REVIEW{Phys. Rev. E}{78}{2008}{011306}.

\bibitem{Staron08}
    \Name{Staron L.}
    \REVIEW{Phys. Rev. E}{77}{2008}{051304}.

\bibitem{Pouliquen08}
     \Name{Pouliquen O. \and Forterre Y.}
     \REVIEW{Annu. Rev. Fluid Mech.}{40}{2008}{124}.

\bibitem{Wyart09}
    \Name{Wyart M.}
    \REVIEW{EPL}{85}{2009}{24003}.

\bibitem{Ertas02}
    \Name{Ertas D. \and Halsey T. C.}
    \REVIEW{Europhys. Lett.}{60}{2002}{931}.

\bibitem{Quere02}
    \Name{Qu\'er\'e D.}
    \REVIEW{Nature Materials}{1}{2002}{14}.

\bibitem{Albert99}
    \Name{Albert R., Pfeifer M. A., Barab\'ask A.-L. \and Schiffer P.}
    \REVIEW{Phys. Rev. Lett.}{82}{1999}{205}.

\bibitem{Seguin11}
    \Name{Seguin A., Bertho Y., Gondret P. \and Crassous J.}
    \REVIEW{Phys. Rev. Lett.}{107}{2011}{048001}.

\bibitem{Courrech05}
    \Name{Courrech du Pont S., Fischer R., Gondret P., Perrin B. \and Rabaud M.}
    \REVIEW{Phys. Rev. Lett.}{94}{2005}{048003}.

\bibitem{Crassous08}
    \Name{Crassous J., M\'etayer J.-F., Richard P. \and Laroche C.}
    \REVIEW{J. Stat. Mech.}{}{2008}{P03009}.

\bibitem{Courrech03}
    \Name{Courrech du Pont S., Gondret P., Perrin B. \and Rabaud M.}
    \REVIEW{Europhys. Lett.}{61}{2003}{492}.

\bibitem{CourrechPhD03}
    \Name{Courrech du Pont S.}
  \Book{PhD Thesis}
  \Year{2003}.

\bibitem{Taberlet03}
    \Name{Taberlet N., Richard P.,Valance A., Losert W., Pasini J.M., Jenkins J.T. \and Delannay R.}
    \REVIEW{Phys. Rev. Lett.}{91}{2003}{264301}.

\end{thebibliography}
\end{document}